\documentclass{jpsj2}
\usepackage{epsf}
\usepackage{graphicx}
\usepackage{multirow}
\usepackage{booktabs}
\usepackage{amsfonts}
\usepackage{dcolumn}%
\usepackage{color,ulem}
\usepackage{bm}
\usepackage{color}

\begin{document}

\title{Theoretical study of magnetism induced by proximity effect in a ferromagnetic Josephson junction with a normal metal}
\author{Shin-ichi Hikino}

\inst{National Institute of Technology. Fukui College, Sabae, Fukui 916-8507, Japan}
\vspace{10pt}
\date{\today}

\abst{
We theoretically study the magnetism induced by the proximity effect in the normal metal of ferromagnetic Josephson junction 
composed of two $s$-wave superconductors separated by ferromagnetic metal/normal metal/ferromagnetic metal junction 
({\it S}/{\it F}/{\it N}/{\it F}/{\it S} junction). 
We calculate the magnetization in the $N$ by solving the Eilenberger equation. 
We show that the magnetization arises in the {\it N} when the product of anomalous Green's functions of 
the spin-triplet even-frequency odd-parity Cooper pair and spin-singlet odd-frequency odd-parity Cooper pair in the {\it N} has a finite value. 
The induced magnetization $M(d,\theta)$ can be decomposed into two parts, $M(d,\theta)=M^{\rm I}(d)+M^{\rm II}(d,\theta)$, where 
$d$ is the thickness of $N$ and $\theta$ is superconducting phase difference between two {\it S}s. 
Therefore, $\theta$ dependence of $M(d,\theta)$ allows us to control the amplitude of magnetization by changing $\theta$. 
The variation of $M(d,\theta)$ with $\theta$ is indeed the good evidence of the magnetization induced by the proximity effect, 
since some methods of magnetization measurement pick up total magnetization in the {\it S}/{\it F}/{\it N}/{\it F}/{\it S} junction. 
}


\maketitle 


\section{Introduction}\label{sec:introduction}
In a superconductor/normal metal ($S$/$N$) junction, 
the pair amplitude of Cooper pair in the $S$ penetrates into the $N$. 
This is called the proximity effect~\cite{degennes}. 
One of crucial phenomena generated by the proximity effect is Josephson effect, which is characterized by 
DC current flowing without a voltage-drop between two {\it S}s separated by normal metal ($N$)~\cite{degennes,likhalev}. 
The Josephson critical current in the $S$/$N$/$S$ junction monotonically decreases with 
increasing the thickness of $N$ ~\cite{degennes,likhalev}. 

The proximity effect in $s$-wave superconductor/ferromagnetic metal ({\it S}/{\it F}) hybrid junctions provides interesting phenomena which are not observed 
in {\it S}/{\it N} hybrid junctions and thus has been extensively studied both theoretically and experimentally~\cite{ryazanov-prl, kontos-prl, sellier-pr, bauer-prl, 
frolov-prb, robinson, born-prb, weides-prl, oboznov-prl, shelukhin-prb, pfeiffer-prb, bannykh-prb, khaire-prb, wild-epjb, kemmer-prb, deutscher, bergeret-prb64,
golubov-rmp, buzdin-rmp, bergeret-rmp}. 
In a {\it S}/{\it F} junction, due to the proximity effect, spin-singlet Cooper pairs (SSCs) penetrate into the {\it F}. 
Because of the exchange splitting of the electronic density of states for up and down electrons, 
the SSC in the F acquires the finite center-of-mass momentum, and the pair amplitude shows a
damped oscillatory behavior with the thickness of the {\it F}~\cite{golubov-rmp,buzdin-rmp,bergeret-rmp}. 
Interesting phenomenon induced by the damped oscillatory behavior of the pair amplitude is a $\pi$-state in 
a {\it S}/{\it F}/{\it S} junction, where the current-phase relation in the Josephson current is shifted by $\pi$ from that of 
the ordinary {\it S}/{\it I}/{\it S} and {\it S}/{\it N}/{\it S} junctions (called 0-state)~\cite{ryazanov-prl, kontos-prl, sellier-pr, bauer-prl, frolov-prb, robinson, 
born-prb, weides-prl, oboznov-prl, shelukhin-prb, pfeiffer-prb, bannykh-prb, khaire-prb, wild-epjb, kemmer-prb, deutscher, bergeret-prb64, 
golubov-rmp, buzdin-rmp, bergeret-rmp}.  

In addition, the proximity effect in {\it S}/{\it F} hybrid junctions also generates 
the spin-triplet odd-frequency even-parity Cooper pairs (STCs) in $F$s, 
although the $S$ is an $s$-wave superconductor~\cite{bergeret-rmp}. 
Here, the anomalous Green's functions of spin-triplet components are odd functions with respect to the Fermion Matsubara 
frequency $\omega_{n}$, i.e., $f({\hat{p}_{\rm F},-\omega_{n} }) = -f({\hat{p}_{\rm F},\omega_{n} })$ and even function 
with respect to the direction of Fermi momentum ${\hat{p}}_{\rm F}={\vec{p}_{\rm F}/|\vec{p}_{\rm F}| } $, i.e., $f(-{\hat{p}_{\rm F},\omega_{n} }) = f({\hat{p}_{\rm F},\omega_{n} })$
($\vec{p}_{\rm F}$ is Fermi momentum). 
It should be noted that the anomalous Green's functions in bulk $s$-wave superconductors are generally even functions 
with respect to $\omega_{n}$ and ${\hat{p}}_{\rm F}$. 
When the magnetization in the {\it F} is uniform in a {\it S}/{\it F} junction, 
the STC composed of opposite spin electrons 
(i.e., total spin projection on $z$ axis being $S_z=0$) and SSCs penetrates into the {\it F} due to the 
proximity effect~\cite{bergeret-rmp,yokoyama-prb75}. 
The penetration length of STC with $S_z=0$ and SSC as described above into the {\it F} is very short and 
the amplitude of STC exhibits a damped oscillatory behavior inside the {\it F} with increasing the thickness of {\it F}. 
The penetration length is determined by $\xi_{\rm F}=\sqrt{\hbar D_{\rm F}/h_{\rm ex}}$, which is typically a order of few nanometers~\cite{ryazanov-prl, kontos-prl, sellier-pr, bauer-prl, frolov-prb, robinson, born-prb, weides-prl, oboznov-prl, shelukhin-prb, pfeiffer-prb, bannykh-prb, khaire-prb, 
wild-epjb, kemmer-prb, golubov-rmp, buzdin-rmp, bergeret-rmp}. 
Here, $D_{\rm F}$ and $h_{\rm ex}$ are the diffusion coefficient and the exchange field in the {\it F}, respectively. 

Moreover, when the magnetization in the {\it F} is non-uniform in a {\it S}/{\it F} junction, 
the STC formed by electrons of equal spin ($|S_z|=1$) can also be induced in the {\it F}. 
This includes cases, for instance, where the $F$ contains a magnetic domain wall~\cite{bergeret-prl86, champel-prb72, braude-prl98, fominov-prb75, 
volkov-prb78, alidoust-prb81, bzdin-prb83}, 
the junction consisting of multilayers of {\it F}s~\cite{volkov-prb90, bergeret-prb68, 
houzet-prb76, trifunovic-prb82, volkov-prb81, trifunovic-prb84, melnikov-prl109, 
knezevic-prb85, richard-prl110, fritsch-njp16, alidoust-prb89, fominov-jetpl77, fominov-jetpl91, kawabata-jpsj82, 
mironov-prb89, halterman-prb94}, or spin active interface at the {\it S}/{\it F} interface
~\cite{eschrig-sh, asano-prl, galaktionov-prb, beri-prb, linder-prb82, trifunovic-prl107, bergeret-prl110, pal-sr7}. 
The penetration length of STC with spin $|S_{z}|=1$ is determined by $\xi_{\rm T}=\sqrt{\hbar D_{\rm F}/2 \pi k_{\rm B} T}$ in $F$s 
($T$ is temperature)~\cite{deutscher}. 
This is approximately 2 orders of magnitude larger than the propagation length of the SSC. 
Thus, the proximity effect of the STC with $|S_{z}|=1$ is called the long-range proximity effect (LRPE).

The LRPE can be observed by the measurement of Josephson critical current in ferromagnetic Josephson junctions with $F$s. 
Because the Josephson critical current carried by STCs with $|S_{z}|=1$ shows monotonically decrease as a function of the thickness of {\it F} and its 
decay length of STC is about determined by $\xi_{\rm T}$. 
The Josephson current carried by the STCs firstly has been predicted theoretically~\cite{bergeret-prl86, eschrig-sh, houzet-prb76} 
and confirmed experimentally~\cite{keizer, robinson-science, khaire, anwar-apl}. 
Instead of measurement of Josephson current, earmark of STC has been observed by measuring the superconducting transition temperature ($T_{\rm C}$)
as a function of polar angle of magnetization between {\it F}1 
and {\it F}2 in spin valves~\cite{leksin-prl109,wang-prl89, singh-prx5}. 

As well as the detection of STC by measuring the Josephson critical current and $T_{\rm C}$ as mentioned above, 
the spin-dependent transport of STC in {\it S}/{\it F} hybrid junctions gives a direct evidence of spin in the STC 
~\cite{bergeret-rmp, Lofwander-prl95, halterman-prb77, alidoust-prb81, pugach-apl101, shomali-njp, hikino-prl110, kukagna-prb90, moor-sst28}. 
One of the simplest quantity induced by the spin of the STC is a magnetization induced by the proximity effect of STC 
~\cite{bergeret-rmp, Lofwander-prl95, alidoust-prb81, pugach-apl101,kukagna-prb90}. 
In F/S junctions, the STC is also induced to {\it S} by the inverse proximity effect\cite{prb-69, prb-72, prb-79, prl-102a, prl-102b}.
The magnetization induced by spin of STC is a good fingerprint to establish the existence of STC. 
{\it F}/{\it S}/{\it F} junctions called spin valve are a typical junction to obtain the magnetization induced by the STC~\cite{bergeret-rmp, Lofwander-prl95, halterman-prb77}. 

To detect the magnetization induced by the STC, recently, 
a rather complex ferromagnetic Josephson junction composed of two {\it S}s separated by {\it F}/{\it F}/{\it S}/{\it F}/{\it F} junction~\cite{pugach-apl101} 
is theoretically proposed  in the diffusive transport region. 
The amplitude of the magnetization in this ferromagnetic Josephson junction can be freely controlled by tuning the superconducting phase difference ($\theta$). 
The variation of magnitude of magnetization by tuning $\theta$ can be detected by using the magnetic sensor based on SQUID~\cite{book-mm}. 
On the other hand, in the ballistic transport region and weak non-magnetic impurity scattering transport region, 
the understanding of magnetic transport in the STC is also important in the ferromagnetic Josephson junction, 
since recent nanofabrication techniques have given the ballistic transport region to ferromagnetic Josephson junctions~\cite{robinson}.

In this paper, we calculate the magnetization induced by the proximity effect in the {\it N} of a simple ferromagnetic Josephson junction composed of two {\it S}s 
separated by {\it F1}/{\it N}/{\it F2} junction 
in ballistic transport region and weak non-magnetic impurity scattering transport region. 
The present {\it S}/{\it F1}/{\it N}/{\it F2}/{\it S} junction is indeed suitable to control the magnetization between two $F$s by using a weak external magnetic field, 
since the N can reduce the exchange coupling between two $F$s\cite{book-mm}.
For this purpose, we solve the Eilenberger equation in the quasiclassical Green's function theory. 
Assuming that the magnetization in $F2$ is fixed along the  $z$ direction perpendicular to the junction direction ($x$ direction), 
we show that i) the $x$ component of the magnetization in the $N$ is always zero, ii)   
the $y$ component becomes exactly zero when the magnetizations in $F$1 and $F$2 are 
collinear, and iii) the $z$ component is generally finite for any magnetization alignment between 
$F$1 and $F$2. 
We find that the magnetization in the $N$ is composed of $\theta$ dependent magnetization and $\theta$ independent magnetization. 
The origin of these magnetizations is due to the spin-triplet even-frequency odd parity Cooper pair 
and spin-singlet odd-frequency odd-parity Cooper pair induced by the proximity effect. 
The magnetization is suppressed with decreasing the non-magnetic impurity scattering time. 
Moreover, the $\theta$ dependence of magnetization disappears with increasing the thickness ($d$) of {\it N}. 
The disappearance of $\theta$ dependence of magnetization is because $\theta$ dependent magnetization shows the rapid decrease with increasing $d$ 
compared with the $\theta$ independent magnetization.

The rest of this paper is organized as follows. 
In Sec.~\ref{sec:formulation}, we introduce a simple {\it S}1/{\it F}1/{\it N}/{\it F}2/{\it S}2 Josephson junction and 
formulate the magnetization induced by the proximity effect in the $N$ of this junction by solving Eilenberger equation in the ballistic transport region 
and the transport region in the case of presence of weak non-magnetic impurity scattering. 
The property of anomalous Green's function in SSC and STC is discussed in the present system. 
In Sec.~\ref{sec:result}, the numerical results of magnetization are given. 
Moreover, the behavior of magnetization with thickness of {\it N} is discussed. 
Finally, the magnetization induced by the proximity effect is estimated for a typical set of realistic parameters in Sec.~\ref{sec:dis}.
The summary of this paper is given in Sec.~\ref{sec:summary}. 

\section{Magnetization in normal metal induced by proximity effect in an S1/F1/N/F2/S2 junction}\label{sec:formulation}
\subsection{Set up of S1/F1/N/F2/S2 junction} 
\begin{figure}[t!]
\begin{center}
\vspace{10 mm} 
\includegraphics[width=8cm]{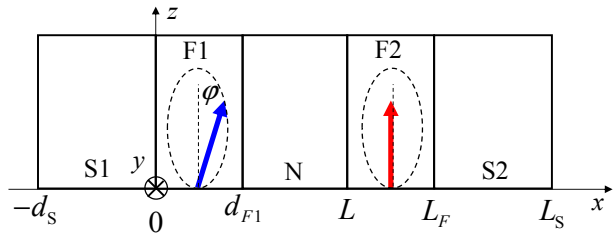}
\caption{ (Color online) 
Schematic illustration of the {\it S}1/{\it F}1/{\it N}/{\it F}2/{\it S}2 junction studied, where the normal metal ({\it N}) is sandwiched by two 
ferromagnetic metals ({\it F}1 and {\it F}2). 
We assume that {\it S}1($S$2) and {\it F}1($F$2) are clean limit and the N is clean limit or weakly disordered metal due to the non-magnetic impurity scattering.  
Arrows in {\it F}1 and {\it F}2 indicate the direction of in-plane ferromagnetic magnetizations. 
While the magnetization in $F$2 is fixed along the $z$ direction, the $F$1 is assumed to be a free layer 
in which the magnetization can be controlled by an external magnetic field within the $yz$ plane 
with $\varphi $ being the polar angle of the magnetization. 
$d_{\rm S}$, $d_{\rm F1}$, $d_{\rm F2}$, and $d$ are the thicknesses of $S$, $F$1, $F$2, and $N$, 
respectively, with $L=d+d_{\rm F1}$, $L_{\rm F}=L+d_{\rm F2}$, and $L_{\rm S}=L_{\rm F}+d_{\rm S}$. 
The uniform magnetizations are assumed in both $F$1 and $F$2 layers. 
}
\label{sfnfs-gm}
\end{center}
\end{figure}
As depicted in Fig.~\ref{sfnfs-gm}, we consider the $S$1/$F$1/$N$/$F$2/$S$2 junction made of normal metal ($N$) 
sandwiched by two layers of ferromagnetic metal ($F$1 and $F$2) 
attached to $s$-wave superconductors ($S$s). 
We assume that the magnetization in $F$2 is fixed along the $z$ direction, 
while the $F$1 is a free layer in which the magnetization can be controlled by an external magnetic field, 
pointing any direction in the $yz$ plane, parallel to the interfaces, with 
$\varphi $ being the polar angle of the magnetization. We also assume that 
the magnetizations in $F$1 and $F$2 are both uniform. 
The thicknesses of $S$, $F$1, $F$2, and $N$ are $d_{\rm S}$, $d_{\rm F1}$, $d_{\rm F2}$, and $d$, 
respectively, with $L=d+d_{\rm F1}$, $L_{\rm F}=L+d_{\rm F2}$, and $L_{\rm S} = L_{\rm F} + d_{\rm S}$. 

\subsection{Anomalous Green's function in normal metal}\label{sec:agf}
To obtain the analytical form of anomalous Green's function, we employ the linearized Eilenberger equation. 
The linearized Eilenberger equation can be applied to $T/T_{C} \approx 1$, since the amplitude of anomalous Green's function becomes small in this case\cite{deutscher}. 
Moreover, it is expected that the amplitude of anomalous Green's function becomes due to the mismatch of the Fermi surfaces between F and S, 
since F has the exchange splitting between up and down Fermi surfaces\cite{bergeret-prb64}. 
Therefore, the linearization of Eilenberger equation can be suitable to the present system. 

In the ballistic transport region and the weak non-magnetic impurity scattering 
transport region, the magnetization inside the N induced by the proximity effect is evaluated by solving 
the linearized Eilenberger equation 
in each region $m$ ($=$ {\it F}1, {\it N}, and {\it F}2)~\cite{buzdin-rmp,golubov-rmp,bergeret-rmp}. 
\begin{eqnarray}
	&&i \hbar \tilde{p}_{\rm F} v_{\rm F} \partial _{x} \hat{f}^{m}(\vec{r} ) +i 2{\rm sgn}(\omega_{n}) \hbar |\omega_{n}| \hat{f}^{m}(\vec{r} )
	-2\hat{\Delta}(x) \nonumber \\
	&&+h_{\rm ex}^{y}(x) \left\{ \sigma_{y}, \hat{f}^{m}(\vec{r} ) \right\} +h_{\rm ex}^{z}(x) \left[\sigma_{z}, \hat{f}^{m}(\vec{r} ) \right], \nonumber \\
	&&+ i {\rm sgn}(\omega_{n}) \frac{\hbar}{\tau} \hat{f}^{m}(\vec{r} )  = \hat{0}, 
\label{eq} 
\end{eqnarray}
where $\vec{r}=(x,\tilde{p}_{\rm F},\omega_{n}) $, $\tilde{p}_{\rm F}=p_{\rm F_{\it x}}/p_{\rm F}=\cos \chi $, 
$\chi$ is the angle between Fermi momentum ($p_{\rm F}$) and $x$ axis, $p_{\rm F_{\it x}}$ is the $x$ component of Fermi momentum, 
and $\omega_{n}$ is the Fermion Matsubara frequency. 
$\tau$ is the non-magnetic impurity scattering time. 
We assume that $\tau$ has a finite value only inside the {\it N}. 
$\hat{f}^{m}(\vec{r} )$ is give by 
\begin{eqnarray}
{\hat f^m}(\vec{r} ) &=& \left( {\begin{array}{*{20}{c}}
	{f_{ \uparrow  \uparrow }^m(\vec{r} )}&{f_{ \uparrow  \downarrow }^m(\vec{r} )}\\
	{f_{ \downarrow  \uparrow }^m(\vec{r} )}&{f_{ \downarrow  \downarrow }^m(\vec{r} )}
	\end{array}} \right), \nonumber \\
	&=& \left( {\begin{array}{*{20}{c}}
	{ - f_{tx}^m(\vec{r} ) + if_{ty}^m(\vec{r} )}&{f_s^m(\vec{r} ) + f_{tz}^m(\vec{r} )}\\
	{ - f_s^m(\vec{r} ) + f_{tz}^m(\vec{r} )}&{f_{tx}^m(\vec{r} ) + if_{ty}^m(\vec{r} )}
\end{array}} \right)
\label{f}. 
\end{eqnarray}
The $s$-wave superconducting gap $\hat{\Delta}(x)$ is finite only in the {\it S} 
and assume to be constant, i.e., 
\begin{eqnarray}
\hat \Delta(x)  = \left\{ \begin{array}{l}
\left( {\begin{array}{*{20}{c}}
0&{ - \Delta_{\rm 1} }\\
\Delta_{\rm 1} &0
\end{array}} \right), - {d_{\rm S}} < x < 0 \\
\left( {\begin{array}{*{20}{c}}
0&{ - \Delta_{\rm 2} }\\
\Delta_{\rm 2} &0
\end{array}} \right), L_{\rm F} < x<L_{\rm S}\\ 
\,\,\,\,\,\,\,\,\,\,\,\,\,\,\,\,\,\,0,\,\,\,\,\,\,\,\,\,\,\,\,\,\,\,\,\,\,\,\,\,\,\,\,\,\,\,\,\,\,{\rm{other}}
\end{array} \right.\ .\nonumber \\
\end{eqnarray}
where $\Delta_{\rm 1(2)} = \Delta e^{i \theta_{\rm 1(2)}}$ ($\Delta$: real) and $\theta_{\rm 1(2)}$ is the superconducting phase in the $S$1 ($S$2) 
(see Fig.~\ref{sfnfs-gm}). 
The exchange field ${\vec h}_{\rm ex}(x)=(h_{\rm ex}^{x}(x),h_{\rm ex}^{y}(x),h_{\rm ex}^{z}(x))$ 
due to the ferromagnetic magnetizations in the {\it F}s is described by 
\begin{eqnarray}
{\vec h_{\rm ex}}\left( x \right) = \left\{ \begin{array}{l}
h_{\rm ex}^{y} {\vec e}_y + h_{\rm ex}^{z} {\vec e}_z\,,\,\,\,\,\,\,\,\,\,0 < x < {d_{F1}}\\
{h_{\rm ex2}}{{\vec e}_z},\,\,\,\,\,\,\,\,\,\,\,\,\,\,\,\,\,\,\,\,\,\,\,\,\,\,\,\,\,L < x < {L_F}\\
0,\,\,\,\,\,\,\,\,\,\,\,\,\,\,\,\,\,\,\,\,\,\,\,\,\,\,\,\,\,\,\,\,\,\,\,\,\,\,\,\,\,\,\,\,\,\,\,\,\,\,\,{\rm{other}}
\end{array}. \right.
\label{exchange}
\end{eqnarray}
where $h_{\rm ex}^y=h_{\rm ex1}\sin \varphi$, $h_{\rm ex}^z=h_{\rm ex1}\cos \varphi$, 
and $\varphi$ is the polar angle of the magnetization in the {\it F}1  (see Fig.~\ref{sfnfs-gm}). 

To obtain the solutions of Eq.~(\ref{eq}), we impose appropriate boundary conditions\cite{demler-prb}, 
\begin{eqnarray}
{\left. {{{\hat f}^{{\rm{S}}}(\vec{r} ) }} \right|_{x = 0}} &=& {\left. {{{\hat f }^{{\rm{F1}}}}(\vec{r} ) } \right|_{x = 0}}
\label{bc1}, \\
{\left. {{{\hat f }^{{\rm{F1}}}(\vec{r} ) }} \right|_{x = {d_{{\rm{F1}}}}}} &=& {\left. {{{\hat f}^{\rm{N}}(\vec{r} )}} \right|_{x = {d_{{\rm{F1}}}}}}
\label{bc2}, \\
{\left. {{{\hat f}^{\rm{N}}(\vec{r} )}} \right|_{x = L}} &=& {\left. {{{\hat f}^{{\rm{F2}}}(\vec{r} ) }} \right|_{x = L}}
\label{bc3}, 
\end{eqnarray}
and 
\begin{eqnarray}
{\left. {{{\hat f}^{{\rm{F2}}}(\vec{r} ) }} \right|_{x = {L_{\rm{F}}}}} &=& {\left. {{{\hat f}^{{\rm{S}}}(\vec{r} ) }} \right|_{x = {L_{\rm{F}}}}}
\label{bc4}. 
\end{eqnarray}
In the present calculation, we apply the rigid boundary condition 
$\sigma_{\rm F}/\sigma_{\rm S} \ll \xi_{\rm F1(2)}/\xi_{\rm S}$ ($\sigma_{\rm S(F)}$ is the conductivity of {\it S }({\it F}) in the case of normal state, 
$\xi_{\rm S}=\hbar v_{\rm F}/2 \pi k_{\rm B} T$ is the superconducting coherence length, and 
$\xi_{\rm F1(2)}=\hbar v_{\rm F}/2h_{\rm ex1(2)} $)~\cite{buzdin-rmp} 
and we assume that $d_{\rm S}$ is much larger than $\xi_{\rm S}$. 
In this case, the anomalous Green's function in the {\it S}1({\it S}2) attached to the {\it F}1 ({\it F}2) can be approximately given by 
\begin{equation}
{\hat f}^{\rm S} (\vec{r}) |_{x=0 (L_{\rm F})} = - \frac{{\hat \sigma}_{y} \Delta  e^{i\theta_{1 (2)}}}{\sqrt{(\hbar \omega)^{2} + \Delta^{2}}}\equiv \hat{F}^{\rm S1(S2)}. 
\end{equation}

Assuming $d_{\rm F1}/\xi_{\rm F1}\ll 1$, we can perform the Taylar expansion of $\hat{f}^{\rm F1}(\vec{r} ) $ with $x$ as follows~\cite{ode-book}, 
\begin{equation}
	\hat{f}(\vec{r} ) \approx \hat{f}^{\rm F1}(d_{\rm F1}, \tilde{p}_{\rm F}, \omega_{n}) + (x-d_{\rm F1}) \partial _{x} \hat{f}^{\rm F1}(d_{\rm F1}, \tilde{p}_{\rm F}, \omega_{n}), 
\label{ff1}
\end{equation}
Applying the boundary condition of Eq.~(\ref{bc1}) to Eq.~(\ref{ff1}) and then substituting Eq.~(\ref{ff1}) into Eq.~(\ref{eq}), 
we can approximately obtain $\hat{f}^{\rm F1}(\vec{r} )$ as 
\begin{eqnarray}
	\hat{f}^{\rm F1}(\vec{r} ) &\approx& \hat{f}^{\rm F1}(d_{\rm F1},\tilde{p}_{\rm F},\omega_{n}) +\frac{x-d_{\rm F1}}{d_{\rm F1}}
	\left[
	\hat{f}^{\rm F1}(d_{\rm F1},\tilde{p}_{\rm F},\omega_{n})
	-
	\hat{F}^{\rm S1}
	\right]
\label{ap-ff1},
\end{eqnarray}
%
where, 
\begin{eqnarray}
	\hat{f}^{\rm F1}(d_{\rm F1},\tilde{p}_{\rm F},\omega_{n})  &=& 
	\left[
	1-{\rm sgn}(\omega_{n}) \frac{2 \hbar |\omega_{n}| d_{\rm F1}}{ \tilde{p}_{\rm F} \hbar v_{\rm F} }
	\right]
	\hat{F}^{\rm S1}
	+
	i \frac{h_{\rm ex}^{y} d_{\rm F1}}{\tilde{p}_{\rm F} \hbar v_{\rm F}}
	\left\{
	\hat{\sigma_{y}}, \hat{F}^{\rm S1} 
	\right\}. 
\label{ff1}
\end{eqnarray}

For $\hat{f}^{\rm F2}(\vec{r})$\cite{note-add1}, by performing the Tayler expansion with $x$ to assume $d_{\rm F2}/\xi_{\rm F2} \ll 1$, 
applying Eq.~(\ref{bc4}) to $\hat{f}^{\rm F2}(\vec{r})$, and then substituting the obtained  $\hat{f}^{\rm F2}(\vec{r})$ into Eq.~(\ref{eq}), 
$\hat{f}^{\rm F2}(\vec{r})$ is approximately given by 
\begin{eqnarray}
	\hat{f}^{\rm F2}(\vec{r}) &\approx & 
		\hat{f}^{\rm F2}(L,\tilde{p}_{\rm F},\omega_{n})
		-\frac{x-L}{d_{\rm F2}}
		\left[
		\hat{f}^{\rm F2}(L,\tilde{p}_{\rm F},\omega_{n})
		-\hat{F}^{\rm S2}
		\right]
\label{ap-ff2}, 
\end{eqnarray}
%
where 
\begin{eqnarray}
	\hat{f}^{\rm F2}(L,\tilde{p}_{\rm F},\omega_{n}) &=&
    \left[
		1
		+
		{\rm sgn}(\omega_{n}) \frac{2 \hbar |\omega_{n}| d_{\rm F2}}{\tilde{p}_{\rm F} \hbar v_{\rm F}} 
	\right]
\hat{F}^{\rm S2}
		-
		i\frac{h_{\rm ex2} d_{\rm F2}}{\tilde{p}_{\rm F} \hbar v_{\rm F}}
		\left[
		\hat{\sigma}_{z}, \hat{F}^{\rm S2}
		\right]. 
\end{eqnarray}

Next, to solve Eq.~(\ref{eq}) in the {\it N}, 
we consider the solution $\hat{f}^{\rm N}(\vec{r})$ of Eq.~(\ref{eq}) as a sum of symmetric ($\hat{f}^{\rm N}_{\rm S}(\vec{r})$) 
and antisymmetric ($\hat{f}^{\rm N}_{\rm A}(\vec{r})$) parts with $\tilde{p}_{\rm F} $ as follows
\begin{equation}
	\hat{f}^{\rm N}(\vec{r}) = \hat{f}^{\rm N}_{\rm A}(\vec{r})+\hat{f}^{\rm N}_{\rm S}(\vec{r})
\label{fn1}. 
\end{equation} 
Substituting Eq.~(\ref{fn1}) into Eq.~(\ref{eq}), we can obtain the equations : 
\begin{eqnarray}
	\hat{f}^{\rm N}_{\rm A} (\vec{r}) &=& - {\rm sgn}(\omega_{n}) \tilde{p}_{\rm F} \frac{ v_{\rm F}}{ 2 |\tilde{\omega}_{n} | } 
		\partial_x \hat{f}^{\rm N}_{\rm S}(\vec{r})
\label{eq-fna}, 
\end{eqnarray}
and 
\begin{eqnarray}
	\partial_{x}^{2}  \hat{f}^{\rm N}_{\rm S}(\vec{r}) - k_{\rm N}^{2} \hat{f}^{\rm N}_{\rm S}(\vec{r}) &=& \hat{0}
\label{eq-fns}, 
\end{eqnarray}
where $k_{\rm N}=2 | \tilde{\omega}_{n} |/v_{\rm F} |\tilde{p}_{\rm F}|$ and $|\tilde{\omega}_{n}| = |\omega_{n}| + 1/2\tau $. 
By solving Eq.~(\ref{eq-fns}) and then using Eq.~(\ref{eq-fna}) to obtain $\hat{f}^{\rm N}_{\rm A}(\vec{r})$, 
the general solution of $\hat{f}^{\rm N}(\vec{r})$ becomes 
\begin{eqnarray}
	\hat{f}^{\rm N}(\vec{r}) &=& 
	\left[
	1-{\rm sgn}(\omega_{n}) \tilde{p}_{\rm F} \frac{v_{\rm F} k_{\rm N}}{ 2|\tilde{\omega}_{n}| }
	\right]
	\hat{A}^{\rm N} e^{k_{\rm N}x} \nonumber \\
	&+&
	\left[
	1+{\rm sgn}(\omega_{n}) \tilde{p}_{\rm F} \frac{v_{\rm F} k_{\rm N}}{ 2|\tilde{\omega}_{n}| }
	\right]
	\hat{B}^{\rm N} e^{-k_{\rm N}x}
\label{fn2}.
\end{eqnarray} 
To determine arbitrary matrix coefficients $\hat{A}$ and $\hat{B}$, we apply Eqs.~(\ref{bc2}) and (\ref{bc3}) to Eq.~(\ref{fn2}). 
As a result, anomalous Green's functions in the {\it N} are found as 
\begin{eqnarray}
	f_{s}^{\rm N} (\vec{r}) &=& f_{1}^{\rm N} (\vec{r}) + f_{2}^{\rm N} (\vec{r})
\label{fs}, \\
	f_{1}^{\rm N} (\vec{r}) &=& 
	-i \frac{\Delta_{\rm 1}}{E_{\omega_{n}} \sinh(k_{\rm N}d)} 
	\sinh[k_{\rm N}(x-L)] \nonumber \\
	&+&
	i \frac{\Delta_{\rm 2}}{E_{\omega_{n}} \sinh(k_{\rm N}d)} 
	\sinh[k_{\rm N}(x-d_{\rm F1})]
\label{fs1}, \\
	f_{2}^{\rm N} (\vec{r}) &=& 
	i \frac{{\rm sgn}(\omega_{n})}{\tilde{p}_{\rm F}}
	\frac{2 \Delta_{\rm 1} |\omega_{n}| d_{\rm F1}}{ v_{\rm F} E_{\omega_{n}} \sinh(k_{\rm N} d) }
	\sinh[k_{\rm N}(x-L)] \nonumber \\
	&+&
	i \frac{{\rm sgn}(\omega_{n})}{\tilde{p}_{\rm F}}
	\frac{2 \Delta_{\rm 2} |\omega_{n}| d_{\rm F2}}{ v_{\rm F} E_{\omega_{n}} \sinh(k_{\rm N} d) }
	\sinh[k_{\rm N}(x-d_{\rm F1})] 
\label{fs2}, \\
	f_{ty}^{\rm N}(\vec{r}) &=& \frac{1}{\tilde{p}_{\rm F}} 
		\frac{2 h_{\rm ex}^{y} d_{\rm F1} \Delta_{\rm 1}}{ \hbar v_{\rm F} E_{\omega_{n}} \sinh(k_{\rm N}d) }
		\sinh[k_{\rm N}(x-L)]
\label{fty}, 
\end{eqnarray}
and 
\begin{eqnarray}
	f_{tz}^{\rm N}(\vec{r}) &=& \frac{1}{\tilde{p}_{\rm F}} 
		\frac{2 h_{\rm ex}^{z} d_{\rm F1} \Delta_{\rm 1}}{ \hbar v_{\rm F} E_{\omega_{n}} \sinh(k_{\rm N}d) }
		\sinh[k_{\rm N}(x-L)] \nonumber \\
		&+&
		\frac{1}{\tilde{p}_{\rm F}} 
		\frac{2 h_{\rm ex2} d_{\rm F2} \Delta_{\rm 2}}{ \hbar v_{\rm F} E_{\omega_{n}} \sinh(k_{\rm N}d) }
		\sinh[k_{\rm N}(x-d_{\rm F1})] 
\label{ftz}. 
\end{eqnarray}
The anomalous Green's function of spin-singlet Cooper pair in Eq.~(\ref{fs}) is given by the sum 
of the spin-singlet even-frequency even-parity Cooper pair given by Eq.~(\ref{fs1}) and 
the spin-singlet odd-frequency odd-parity Cooper pair given by  Eq.~(\ref{fs2}). 
It should be noticed that the spin-singlet odd-frequency odd-parity Cooper pair in Eq.~(\ref{fs2}) contributes 
to the magnetization ($M_{y(z)}(d,\theta)$) induced by the proximity effect, 
since $M_{y(z)}(d,\theta) \propto {\rm sgn}(\omega_{n}) {\rm Im} [f_{s}^{\rm N}(\vec{r}) f_{ty(tz)}^{\rm N}(\vec{r}) ]$ (see Eq.~(\ref{m})). 
Moreover, it is immediately found that Eqs.~(\ref{fty}) and (\ref{ftz}) indicate the spin-triplet even-frequency odd-parity Cooper pair. 
These results are summarized in Table I. 
The general classification of symmetry of anomalous Green's functions is given by Ref[78]. 
Here, it should be noted that $f_{tx}^{\rm N}(x)=0$ because the exchange field in the $F$1 does not 
have the $x$ component. 
Therefore, the $x$ component of the magnetization in the $N$ is always zero, as discussed below. 

\begin{table}[htb]
\caption{The symmetry of anomalous Green's function given by Eqs.~(\ref{fs})-(\ref{ftz}) in the $N$ of the {\it S}1/{\it F}1/{\it N}/{\it F}2/{\it S}2 junction. 
Where $\omega_{n}$ is the Fermion Matsubara frequency and $\tilde{p}_{\rm F}=\cos\chi$. 
$\chi$ is the angle between the Fermi momentum and the $x$ axis.}
\scalebox{1.4}[1.4]{ 
	\begin{tabular}{|c|c|c|} \hline
	Spin & Frequency $(\omega_{n})$ & Parity $(\tilde{p}_{\rm F})$ \\ \hline
\multirow{2}{*}{Singlet} 
		& Even & Even \\ \cline{2-3} 
			   & Odd & Odd \\ \hline
	Triplet & Even & Odd \\ \hline
	\end{tabular}
	}
\end{table}

\subsection{Magnetization in a normal metal}

Within the quasiclassical Green's function theory, 
the magnetization $M_{y(z)}(d,\theta)$ ($\theta$~:~superconducting phase difference between {\it S}1 and {\it S}2) induced inside the {\it N} is given by~\cite{Lofwander-prl95,champel-prb72} 
\begin{eqnarray}
	\vec{M}(d,\theta)  &=&
		(M_{x}(d,\theta),M_{y}(d,\theta),M_{z}(d,\theta)) \nonumber \\
		&=&
		\frac{A}{V}
		\int_{d_{F1}}^{L}
		\vec{m} (x,\theta) dx
\label{md}, 
\end{eqnarray}
where $\theta=\theta_{\rm 2}-\theta_{\rm 1}$ is the superconducting phase difference between $S2$ and $S1$, 
\begin{eqnarray}
	\vec{m} (x,\theta) &=& (m_{x}(x,\theta),m_{y}(x,\theta),m_{z}(x,\theta)) \nonumber \\
	&=& \frac{g \mu_{\rm B} \pi N_{\rm F} k_{\rm B} T }{2}
	\sum_{\omega_{n}}
	{\rm sgn}(\omega_{n})
	\int_{-1}^{1} d\tilde{p}_{\rm F}
	{\rm Im}
	\left[
	f_{s}^{\rm N}(\vec{r}) \vec{f}_{t}^{\rm N,*}(\vec{r})
	\right]
\label{m},
\end{eqnarray}
and 
\begin{equation}
{\vec{f}_{t}}^{\rm N}(x) = 
	(f_{tx}^{\rm N}(\vec{r}),-f_{ty}^{\rm N}(\vec{r}),f_{tz}^{\rm N}(\vec{r}))
\label{vec-f}. 
\end{equation}
Here, $\vec m(x,\theta)$ is the local magnetization density in the $N$ and $g$ is the $g$ factor of electron, $\mu_{\rm B}$ is the Bohr magneton. 
$A$ and $V=Ad$ are the cross-section area of junction and the volume of $N$, respectively. 
In the quasiclassical Green's function theory, the density of states $N_{\rm F}$ per unit volume 
and per electron spin at the Fermi energy for up and down electrons in the spin polarized $N$ due to 
the proximity effect is assumed to be approximately the same~\cite{golubov-rmp,buzdin-rmp,bergeret-rmp}. 
It should be noticed in Eq.~(\ref{m}) that $m_{x}(x,\theta)$ is always zero because 
$f_{tx}^{\rm N}(x)=0$ (see Sec.~\ref{sec:agf}) and thus the $x$ component $M_{x} (d,\theta)$ 
of the magnetization is always zero. 
Therefore, in the following, we only consider the $y$ and $z$ components of the magnetization. 

Substituting Eqs~(\ref{fs})-(\ref{ftz}) into Eq.~(\ref{m}) and then performing the integration with respect to $x$ in Eq.~(\ref{md}), 
we can obtain the $y$ and $z$ components of magnetization in the $N$ induced by the proximity effect. 
The $y$ component of magnetization in the {\it N} is given as 
\begin{eqnarray}
	M_{y}(d,\theta) &=& M_{y}^{\rm I}(d) + M_{y}^{\rm II}(d,\theta)
\label{my},
\end{eqnarray}
where 
\begin{eqnarray}
	M_{y}^{\rm I}(d) &=& 8g{\mu _{\rm B}}\pi {N_{\rm F}}{\Delta ^2}\frac{{h_{\rm ex}^yd_{\rm F1}^2}}{{{{\left( {\hbar {v_{\rm F}}} \right)}^2}\beta }}
		\sum\limits_{{\omega _n} > 0} {\int_0^1 {d\tilde{p}_{\rm F} } } {K_{{\omega _n}}}\left( {d,\tilde{p}_{\rm F} } \right)
\label{my1}, 
\end{eqnarray}
and 
\begin{eqnarray}
	M_{y}^{\rm II}(d,\theta) &=& 8g{\mu _B}\pi {N_F}{\Delta ^2}\frac{{h_{\rm ex}^y{d_{F1}}{d_{F2}}}}{{{{\left( {\hbar {v_F}} \right)}^2}\beta }}
		\sum\limits_{{\omega _n} > 0} {\int_0^1 {d\tilde{p}_{\rm F} } {R_{{\omega _n}}}\left( {d,\tilde{p}_{\rm F} } \right)\cos \theta }
\label{my2}. 
\end{eqnarray}
Here, we have introduced 
\begin{eqnarray}
	K_{\omega_{n}} (d,\mu) &=& \frac{ \hbar |\omega_{n}| }{ [\tilde{p}_{\rm F} E_{\omega_{n}} \sinh(k_{\rm N}d) ]^2 }
			\frac{ \sinh(2k_{\rm N}d) - 2k_{\rm N}d }{4k_{\rm N}d} \nonumber \\
\end{eqnarray}
and 
\begin{eqnarray}
	R_{\omega_{n}} (d,\mu) &=& \frac{ \hbar |\omega_{n}| }{ [\tilde{p}_{\rm F} E_{\omega_{n}} \sinh(k_{\rm N}d) ]^2 }
			\frac{\sinh(k_{\rm N}d) - k_{\rm N}d \cosh(k_{\rm N}d)}{ 2k_{\rm N}d }. \nonumber \\
\end{eqnarray}
From Eqs.~(\ref{my1}) and (\ref{my2}), it is immediately found 
that $y$ components $M_{y}^{\rm I}(d)$ and $M_{y}^{\rm II}(d,\theta)$ of magnetization is always zero for $\varphi =0$ or $\pi$, 
since $M_{y}^{\rm I}(d)$ and $M_{y}^{\rm II}(d,\theta)$ is proportional to $h_{\rm ex}^{y}=h_{\rm ex1} \sin\varphi$. 
Similarly, the $z$ component of magnetization in the {\it N} is decomposed into two parts,   
\begin{eqnarray}
	M_{z}(d,\theta) &=& M_{z}^{\rm I}(d) + M_{z}^{\rm II}(d,\theta)
\label{mz},
\end{eqnarray}
where 
\begin{eqnarray}
	M_{z}^{\rm I}(d) &=& - 8g{\mu_{\rm B}}\pi {N_{\rm F}}\frac{{{\Delta ^2}}}{\beta (\hbar v_{\rm F})^{2} }\left[ 
		h_{\rm ex}^{z} d_{\rm F1}^{2} + h_{\rm ex2} d_{\rm F2}^{2}
		\right]
		\sum\limits_{{\omega _n} > 0} {\int_0^1 {d\tilde{p}_{\rm F} } {K_{{\omega _n}}}\left( {d,\tilde{p}_{\rm F} } \right)}
\label{mz1}, 
\end{eqnarray}
and 
\begin{eqnarray}
	M_{z}^{\rm II}(d,\theta) &=& - 8g{\mu_{\rm B}}\pi {N_{\rm F}}\frac{{{\Delta ^2}}}{\beta }
		\frac{{{d_{F1}}}}{{\hbar {v_{\rm F}}}}\frac{{{d_{\rm F2}}}}{{\hbar {v_{\rm F}}}}
		\left( h_{\rm ex}^z + h_{\rm ex2}
		\right)
		\sum\limits_{{\omega _n} > 0} {\int_0^1 {d\tilde{p}_{\rm F} } } {R_{{\omega _n}}}\left( {d,\tilde{p}_{\rm F} } \right)\cos \theta
\label{mz2}, 
\end{eqnarray}
It is immediately found that from Eq.~(\ref{mz1}) $M_{z}^{\rm I}(d)$ is always zero when $h_{\rm ex}^{z}d_{\rm F1}^{2}=-h_{\rm ex2}d_{\rm F2}^{2}$, 
whereas from Eq.~(\ref{mz2}), $M_{z}^{\rm II}(d,\theta)$ is always zero when $h_{\rm ex}^{z}=-h_{\rm ex2}$. 

\section{Results}\label{sec:result}
\subsection{Magnetization-phase relation}
\begin{figure}[b!]
\begin{center}
\vspace{30 mm} 
\includegraphics[width=0.8\hsize]{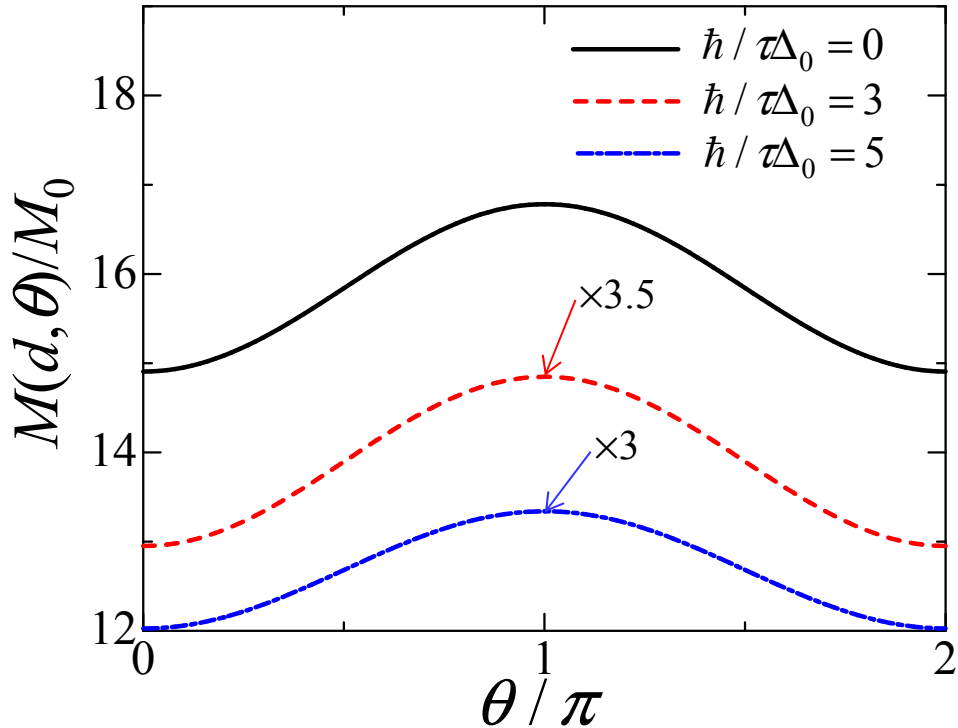}
\caption{ (Color online) Magnetization induced by the proximity effect in the {\it N} 
as a function of $\theta$ for different $\hbar/\tau \Delta_{0}$ indicated in the figure. 
Here we set $T/T_{\rm C} = 0.3$, $\varphi =\pi/2$, $d/\xi_{0}=0.5$, $d_{\rm F1}/\xi_{0} = 0.3$, $d_{\rm F2}/\xi_{0} = 0.2$, $h_{\rm ex1}/\Delta_{0}=30$, 
$h_{\rm ex2}/\Delta_{0}=20$. 
$\xi_{0}=\hbar v_{\rm F}/2 \pi k_{\rm B} T_{\rm C}$ and $M_{0}=g \mu_{\rm B} N_{\rm F} \Delta_{0}$. 
}
\label{m-q-t}
\end{center}
\end{figure}

Let us first numerically evaluate the amplitude of the magnetization in the {\it N} by using Eqs.~(\ref{my}) and (\ref{mz}), i.e., 
\begin{equation}
M(d,\theta)=\sqrt{ [M_{y}(d,\theta)]^{2} + [M_{z}(d,\theta)]^{2} }
\label{mt}.
\end{equation}
In order to perform the numerical calculation of $M(d,\theta)$, 
the temperature dependence of $\Delta$ is assumed to be $\Delta=\Delta_{0}\tanh(1.74\sqrt{T_{\rm C}/T-1})$, 
where $\Delta_{0}$ is the superconducting gap at zero temperature. 
The thickness of $N$, $F1$, and $F2$ is normalized by $\xi_{0}=\hbar v_{\rm F}/2\pi k_{\rm B} T_{\rm C}$. 
Figure~\ref{m-q-t} shows $M(d,\theta)$ normalized by $M_{0}=g \mu_{\rm B} N_{\rm F} \Delta_{0}$ as a function of $\theta$ for different $\hbar/\tau$. 
It is found that $M(d,\theta)$ decreases 
with increasing $\hbar/\tau$ as shown in solid (black), dashed (red), and chain (blue) lines of Fig.~\ref{m-q-t}. 
This reduction of $M(d,\theta)$ is due to the suppression of pair amplitude caused by the non-magnetic impurity scattering, i.e., finite $\hbar/\tau$ inside the {\it N}. 
The variation of $M(d,\theta)$ with respect to $\theta$ is the good fingerprint of magnetization induced by the proximity effect, 
since some experimental methods of magnetization measurement pick up magnetization of {\it F}s to be constant and $M(d,\theta)$. 
Moreover, it should be noticed that $M(d,\theta)$ has a periodicity of $2 \pi$ as expressed by Eqs.~(\ref{my2}) and (\ref{mz2}). 

\begin{figure}[t!]
\begin{center}
\vspace{15 mm} 
\includegraphics[width=0.8\hsize]{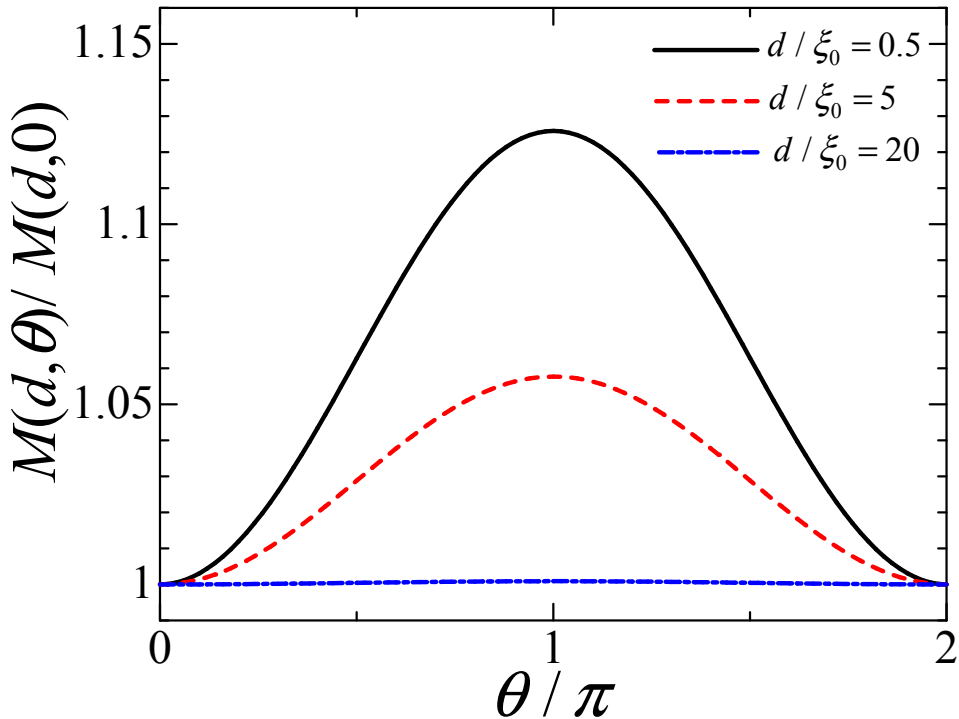}
\caption{ (Color online) Magnetization induced by the proximity effect in the {\it N} 
as a function of $\theta$ for different $d/\xi_{0}$ indicated in the figure. 
Here we set $T/T_{\rm C} = 0.3$, $\varphi =\pi/2$, $\hbar/\tau \Delta_{0}=0$, $d_{\rm F1}/\xi_{0} = 0.3$, $d_{\rm F2}/\xi_{0} = 0.2$, $h_{\rm ex1}/\Delta_{0}=30$, 
$h_{\rm ex2}/\Delta_{0}=20$. 
$\xi_{0}=\hbar v_{\rm F}/2 \pi k_{\rm B} T_{\rm C}$. 
}
\label{m-q-d}
\end{center}
\end{figure}

Figure~\ref{m-q-d} shows the representative result of $M(d,\theta)$ normalized by $M(d,0)$ as a function of $\theta$ for different thickness $d$ of {\it N}. 
From Fig.~\ref{m-q-d}, it is immediately found that $\theta$ dependence of $M(d,\theta)$ gradually vanishes away with increasing $d$. 
For the thin thickness of {\it N} as shown in the solid (black) line of Fig.~\ref{m-q-d}, 
The amplitude of $M(d,\theta)$ exhibits the clear modulation as a function of $\theta$. 
With increasing $d$ as shown in the dashed (red) line of Fig.~\ref{m-q-d}, 
the amplitude of $M(d,\theta)$ decreases but still we can find the modulation of $M(d,\theta)$ with respect to $\theta$. 
However, for the further increase of $d$ as shown in the chain (blue) line of Fig.~\ref{m-q-d}, 
the modulation of $M(d,\theta)$ with respect to $\theta$ is no longer acquired. 
Therefore, from these results, it is found that $d$ is an important parameter to obtain $M(d,\theta)$ controlled by changing $\theta$. 
We will discuss $d$ dependence of magnetization in the next subsection. 

\subsection{Thickness dependence of magnetization in normal metal}
\begin{figure}[t!]
\begin{center}
\vspace{10 mm} 
\includegraphics[width=6.5cm]{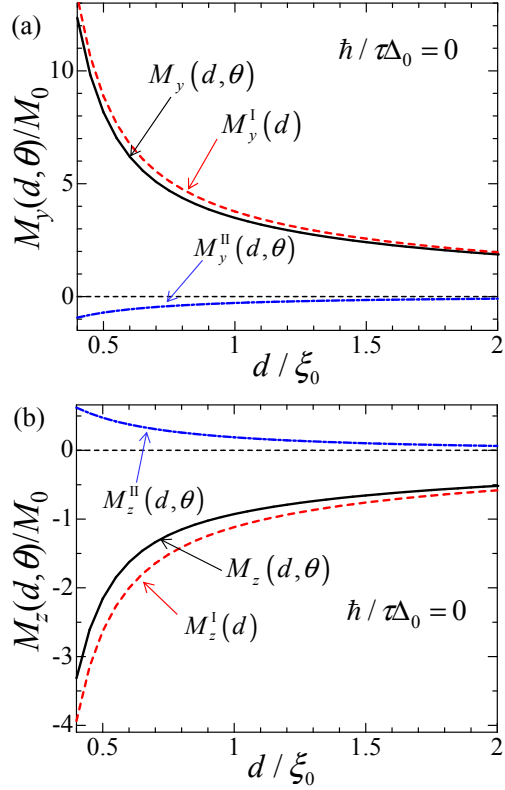}
\caption{ (Color online) The magnetization of $y$ component $M_{y}(d,\theta)$ (a) and $z$ component $M_{z}(d,\theta)$ (b) as a function of $d$. 
Here we set $T/T_{\rm C} = 0.3$, $\theta=0$, $\varphi =\pi/2$, $\hbar/\tau \Delta_{0}=0$, 
$d_{\rm F1}/\xi_{0} = 0.3$, $d_{\rm F2}/\xi_{0} = 0.2$, $h_{\rm ex1}/\Delta_{0}=30$, 
$h_{\rm ex2}/\Delta_{0}=20$. 
$\xi_{0}=\hbar v_{\rm F}/2 \pi k_{\rm B} T_{\rm C}$ and $M_{0}=g \mu_{\rm B} N_{\rm F} \Delta_{0}$. 
We also plot $M_{y(z)}^{\rm I}(d)$ and $M_{y(z)}^{\rm II}(d,\theta)$, separately. 
}
\label{md1}
\end{center}
\end{figure}

\begin{figure}[t!]
\begin{center}
\vspace{15 mm} 
\includegraphics[width=6.5cm]{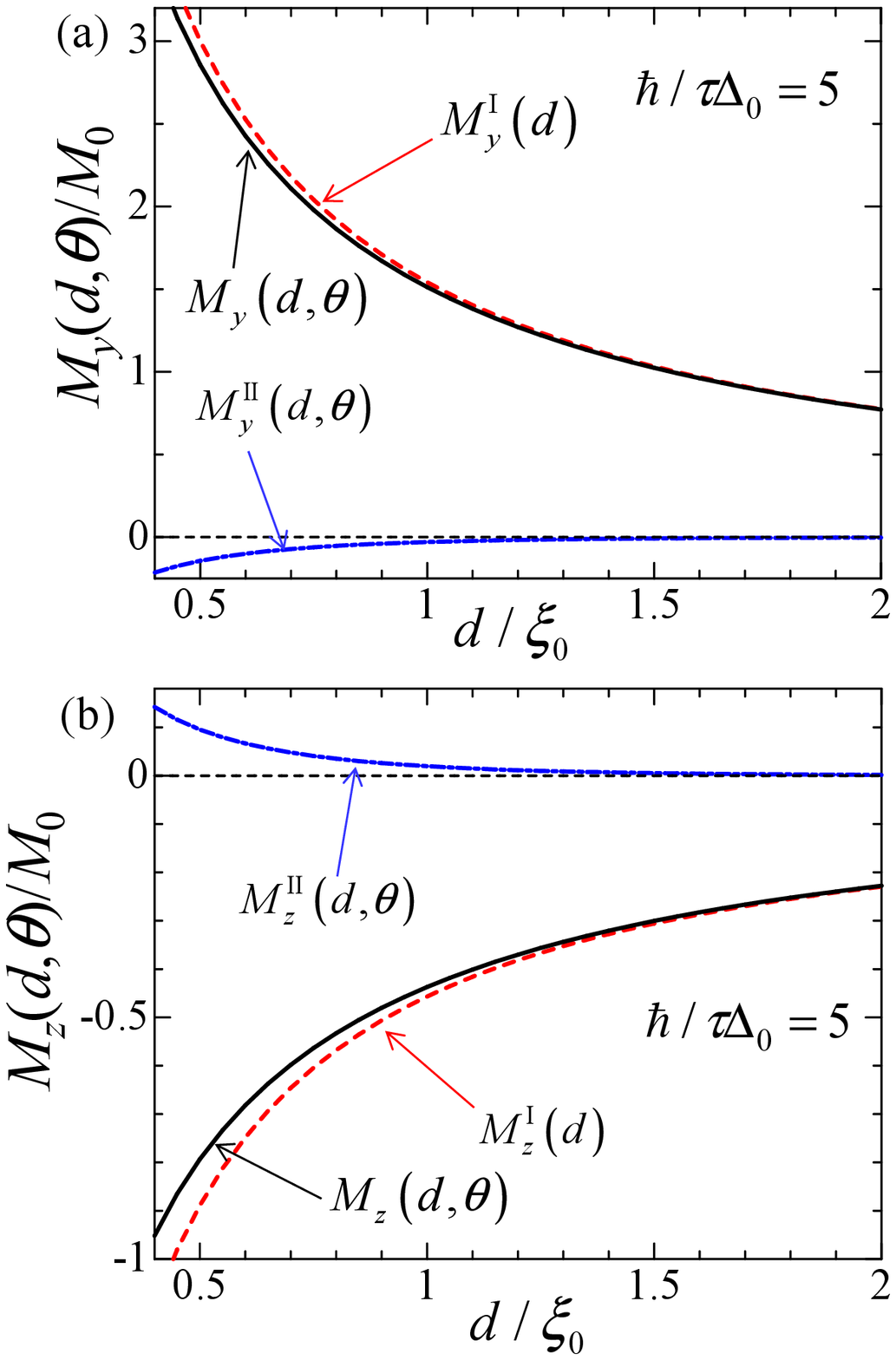}
\caption{ (Color online) The magnetization of $y$ component $M_{y}(d,\theta)$ (a) and $z$ component $M_{z}(d,\theta)$ (b) as a function of $d$. 
Here we set $T/T_{\rm C} = 0.3$, $\theta=0$, $\varphi =\pi/2$, $\hbar/\tau \Delta_{0}=5$, 
$d_{\rm F1}/\xi_{0} = 0.3$, $d_{\rm F2}/\xi_{0} = 0.2$, $h_{\rm ex1}/\Delta_{0}=30$, 
$h_{\rm ex2}/\Delta_{0}=20$. 
$\xi_{0}=\hbar v_{\rm F}/2 \pi k_{\rm B} T_{\rm C}$ and $M_{0}=g \mu_{\rm B} N_{\rm F} \Delta_{0}$. 
We also plot $M_{y(z)}^{\rm I}(d)$ and $M_{y(z)}^{\rm II}(d,\theta)$, separately. 
}
\label{md2}
\end{center}
\end{figure}
Let us now evaluate numerically the $d$ dependence of magnetization induced by the proximity effect in the {\it N} by using Eqs.~(\ref{my}) and (\ref{mz}). 
Figure~\ref{md1} shows the magnetization $M_{y(z)}(d,\theta)$ as a function of $d$ for $\hbar/\tau=0$. 
Here, it should be noticed that we separately plot $y$ and $z$ component of magnetization ($M_{y}(d,\theta)$ 
and $M_{z}(d,\theta)$) as a function of $d$. 
At the same moment, $M_{y(z)}^{\rm I}(d)$ and $M_{y(z)}^{\rm II}(d,\theta)$, which are plotted by dashed (red) and chain (blue) lines are also indicated in Fig.~\ref{md1}. 
Figure~\ref{md1} shows $M_{y}(d,\theta)$ and $M_{z}(d,\theta)$ as a function of $d$ for $\hbar/\tau \Delta_{0}=0$ and $\varphi =\pi/2$. 
From Figs.~\ref{md1} (a) and (b), we find that $M_{y}(d,\theta)$ and $M_{z}(d,\theta)$ exhibits monotonically decrease with increasing $d$. 
It is also found that the damping rate of $M_{y(z)}^{\rm I}(d)$ with $d$ is remarkably weak compared with that of $M_{y(z)}^{\rm II}(d,\theta)$ 
as shown in dashed (red) and chain (blue) lines of Figs.~\ref{md1} (a) and (b). 
Therefore, when $d/\xi_{0}\gg 1$, the main contribution to $M_{y(z)}(d,\theta)$ arises from $M_{y(z)}^{\rm I}(d)$, 
since $M_{y(z)}^{\rm II}(d,\theta)$ is vanishingly small. 
This is the reason why the modulation of magnetization with $\theta$ disappears for $d/\xi_{0}\gg 1$ as shown in Fig.~\ref{m-q-d}. 
Moreover, it is immediately realized that the sign of $M_{y(z)}^{\rm I}(d)$ and $M_{y(z)}^{\rm II}(d,\theta)$ is different for any $d$ from Fig.~\ref{md1} (a) and (b). 
Figure~\ref{md2} shows $M_{y}(d,\theta)$ and $M_{z}(d,\theta)$ as a function of $d$ for $\hbar/\tau \Delta_{0}=5$ and $\varphi =\pi/2$. 
As compared to Fig.~\ref{md1} (a) and (b), Fig.~\ref{md2} (a) and (b) indicate that the magnitude of magnetization is suppressed 
and the damping rate of magnetization with $d$ is stronger as $\tau$ decreases. 
For $\hbar/\tau \Delta_{0}=5$, $M_{y(z)}^{\rm II}(d,\theta)$ is ignorable small around $d/\xi_{0}=1.5$. 
Therefore, $M_{y(z)}(d,\theta)$ almost becomes independent with $\theta$ for $d/\xi_{0}>1.5$.

\section{Discussions}\label{sec:dis} 
Here, we shall discuss the $d$ dependence of $M_{y}(d,\theta)$ and $M_{z}(d,\theta)$ by using approximated formula
of Eqs.~(\ref{my1})-(\ref{mz2}).
For $T\approx T_{\rm C}$, $\hbar/\tau=0$, and  $d/\xi_{0}\gg 1$, the $y$ components $M_{y}^{\rm I}(d)$ and $M_{y}^{\rm II}(d,\theta)$ 
of magnetization are approximately given as 
\begin{eqnarray}
	M_{y}^{\rm I}(d) &\approx& 4 g \mu \pi N_{\rm F} \Delta^{2}
		\frac{ d_{\rm F1}^{2} k_{\rm B} T }{ (\hbar v_{\rm F})^{2} } h_{\rm ex}^{y} \frac{\xi_{0}}{d}
\label{my1-ap}
\end{eqnarray}
and 
\begin{eqnarray}
	M_{y}^{\rm II}(d,\theta) &\approx& -4 g \mu \pi N_{\rm F} \Delta^{2}
		\frac{ d_{\rm F1} d_{\rm F2} k_{\rm B} T }{ (\hbar v_{\rm F})^{2} } \nonumber \\
			&\times & 
			h_{\rm ex}^{y}
			\left(
			1-\frac{2 \xi_{0}}{d} 
			\right)\exp(-d/\xi_{0}) \cos\theta
\label{my2-ap}, 
\end{eqnarray}
whereas the $z$ components $M_{z}^{\rm I}(d)$ and $M_{z}(d,\theta)$ of magnetization are approximately given as 
\begin{eqnarray}
	M_{z}^{\rm I}(d) &\approx & - \frac{4 g \mu \pi N_{\rm F} \Delta^{2} k_{\rm B} T}{ (\hbar v_{\rm F})^{2} } \nonumber \\
		&\times&
		\left[\
		h_{\rm ex}^{z} 
		\left(  
		d_{\rm F1} 
		\right)^{2}
		+
		h_{\rm ex2} 
		\left(
		d_{\rm F2}
		\right)^{2}
		\right]\	
		\frac{\xi_{0}}{d}
\label{mz1-ap} 
\end{eqnarray}
and 
\begin{eqnarray}
	M_{z}^{\rm II}(d,\theta) &\approx & 4 g \mu \pi N_{\rm F} \Delta^{2} k_{\rm B} T 
	\frac{ d_{\rm F1} d_{\rm F2} }{ (\hbar v_{\rm F})^{2} }  \nonumber \\
	&\times &
	(h_{\rm ex}^{z} + h_{\rm ex})
	\left(
	1-\frac{2 \xi_{0}}{d} 
	\right)\exp(-d/\xi_{\rm 0}) \cos\theta
\label{mz2-ap}. \nonumber \\
\end{eqnarray}
From Eqs.~(\ref{my1-ap})-(\ref{mz2-ap}), it is found that $M_{y(z)}^{\rm I}(d)$ algebraically decreases as $1/d$, 
whereas  $M_{y(z)}^{\rm II}(d,\theta)$ exponentially decreases with $d$. 
Moreover, from Eqs.~(\ref{my1-ap})-(\ref{mz2-ap}), it is also found that the sign of $M_{y(z)}^{\rm I}(d)$ and $M_{y(z)}^{\rm II}(d,\theta)$ is always opposite. 
Here, it should be noticed that the coefficient (1-2$\xi_{0}/d$) of Eqs.~(\ref{my2-ap}) and (\ref{mz2-ap}) is always positive value, 
since $\xi_{\rm 0}/d$ is much smaller than 1 in the present approximation. 
These results are indeed qualitatively consistent with numerical results as shown in Fig.~\ref{md1} and \ref{md2}.

Finally, we shall approximately estimate $\xi_{0}$ and the amplitude of magnetization in the {\it N}. 
In clean normal metals, $\xi_{0}$ is in a range of several hundred nanometers~\cite{deutscher}. 
Therefore, the magnetization in the {\it N} induced by the proximity effect has a finite value in this length scale. 
The amplitude $M$ of the magnetization is estimated to be of order 
$g \mu_{\rm B} N_{\rm F} \Delta_{0}$ (see Fig.~\ref{md1} and \ref{md2}). 
When we use a typical set of parameters\cite{note7,book-ssp}, the amplitude $M$ is approximately 100 $\rm{A/m}$. 
It is expected that this value can be detected by the magnetization measurement by utilizing SQUID~\cite{book-mm}. 

\section{Summary}\label{sec:summary}
We have calculated the magnetization induced by the proximity effect in the {\it N} of the {\it S}1/{\it F}1/{\it N}/{\it F}2/{\it S}2 Josephson junction. 
Where it has been assumed that the magnetization of $F1$ is in the $yz$ plane and the magnetization of $F2$ is fixed along with $z$ axis. 
Based on the quasiclassical Green's function theory in the ballistic transport region 
and the weak non-magnetic impurity scattering region, 
we have found that the magnetization in the $N$ are induced by the emergence of the spin-triplet even-frequency odd-parity Cooper pair 
and spin-singlet odd frequency odd parity Cooper pair, which are induced by the proximity effect in the {\it S}1/{\it F}1/{\it N}/{\it F}2/{\it S}2 junction. 
We have shown that i) the $x$ component of the magnetization in the $N$ is always zero, 
ii) the $y$ component is exactly zero
when the magnetization direction between $F$1 and $F$2 is collinear, 
and iii) the $z$ component is generally finite for any magnetization direction between $F$1 and $F$2. 

We have found that the magnetization in the {\it N} can be controlled by tuning $\theta$. 
This magnetization is suppressed with decreasing the relaxation time of the non-magnetic impurity scattering. 
Moreover, it has been found that $\theta$ dependence of magnetization vanishes away when thickness $d$ of {\it N} is much larger than $\xi_{0}$. 

The magnetization induced in the $N$ can be decomposed into $\theta$ dependent and $\theta$ independent parts. 
We have shown that $\theta$ dependent magnetization rapidly decays with increasing $d$, 
whereas  $\theta$ independent magnetization slowly decays with increasing the thickness of $N$. 
We have also found that the sign of $\theta$ independent and $\theta$ depend magnetizations are always opposite. 
This $\theta$ dependence of magnetization is important to confirm the existence of spin-triplet Cooper pair, 
since some experimental methods of magnetization measurement pick up net magnetization in the present ferromagnetic Josephson junction.

\section*{Acknowledgments} 
The authors would like to thank S. Yunoki for useful discussions and comments.

{}

\begin{thebibliography}{}
\bibitem{note-add1}
Eq.(\ref{ff1}) becomes the anomalous Green's function of normal metal when the exchange field is equal to zero, 
i.e., the second term of Eq.(\ref{ff1}) vanishes.
\bibitem{note7}
$N_{\rm F}=\frac{1}{4\pi^2}\left( \frac{2m}{\hbar^2} \right)^{3/2}\varepsilon_{\rm F}^{1/2} \approx 5.4 \times10^{27}\,\ {\rm eV^{-1}} {\rm m^{-3}} $ 
per spin with the Fermi energy 
$\varepsilon_{\rm F}\approx5$ eV~\cite{book-ssp}, where $m$ is the electron mass. 
$\Delta_{0}=1$ meV for Nb~\cite{deutscher}.
\end{thebibliography}
\end{document}